\documentclass[useAMS,usenatbib]{mn2e}

\usepackage{graphicx}
\usepackage{subfigure}
\usepackage{amssymb}

\newcommand{\Msun}{M_\odot}

\newcommand{\Rsun}{R_\odot}

\title[Simulating protostellar evolution]{Simulating protostellar evolution and radiative feedback in the cluster environment}
\author[M.\ Klassen, R.E.\ Pudritz, \& T.\ Peters ]{Mikhail Klassen$^{1}$\thanks{E-mail: klassm@mcmaster.ca}, Ralph E.\ Pudritz$^{1,2}$, Thomas Peters$^{3,4}$\\
$^{1}$Department of Physics and Astronomy, McMaster University, 1280 Main St.~W, Hamilton, ON L8S 4M1, Canada\\
$^{2}$Origins Institute, McMaster University, 1280 Main St.~W, Hamilton, ON L8S 4M1, Canada\\
$^{3}$Zentrum f\"{u}r Astronomie der Universit\"{a}t Heidelberg, Institut f\"{u}r Theoretische Astrophysik, Albert-Ueberle-Str. 2, D-69120 Heidelberg, \\Germany\\
$^{4}$Institut f\"{u}r Theoretische Physik, Universit\"{a}t Z\"{u}rich, Winterthurerstrasse 190, CH-8057 Z\"{u}rich, Switzerland}

\begin{document}
\bibliographystyle{mn2e}

\date{15 December 2011}

\pagerange{\pageref{firstpage}--\pageref{lastpage}} \pubyear{2011}

\maketitle

\label{firstpage}

\begin{abstract}
Radiative feedback is among the most important consequences of clustered star formation inside molecular clouds. At the onset of star formation, radiation from massive stars heats the surrounding gas, which suppresses the formation of many low-mass stars. When simulating pre-main-sequence stars, their stellar properties must be defined by a prestellar model. Different approaches to prestellar modeling may yield quantitatively different results. In this paper, we compare two existing prestellar models under identical initial conditions to gauge whether the choice of model has any significant effects on the final population of stars. The first model treats stellar radii and luminosities with a ZAMS model, while separately estimating the accretion luminosity by interpolating to published prestellar tracks. The second, more accurate prestellar model self-consistently evolves the radius and luminosity of each star under highly variable accretion conditions. Each is coupled to a raytracing-based radiative feedback code that also treats ionization. The impact of the self-consistent model is less ionizing radiation and less heating during the early stages of star formation. This may affect final mass distributions. We noted a peak stellar mass reduced by 8\% from 47.3$\Msun$ to 43.5$\Msun$ in the evolutionary model, relative to the track-fit model. Also, the difference in mass between the two largest stars in each case is reduced from 14$\Msun$ to 7.5$\Msun$. The HII regions produced by these massive stars were also seen to flicker on timescales down to the limit imposed by our timestep ($<$ 560 years), rapidly changing in size and shape, confirming previous cluster simulations using ZAMS-based estimates for prestellar ionizing flux.
\end{abstract}


\section{Introduction}

The conversion of molecular gas into fully-formed stars is complex, involving several diverse processes. These different processes are linked to each other through feedback mechanisms that make isolating and understanding the contribution of each process a difficult task. A key point in this regards is that stars also rarely form in isolation, but instead are seen to be forming in clusters and subclusters within molecular clouds \citep{Clarke+2000,Testi+2000}. In the cluster environment, the formation of a sufficiently massive star can affect all the others through the energy it radiates back into the cloud. Numerical simulations of star formation have made it very clear that the effects of stellar radiation cannot be neglected. Simulations including some form of radiative transfer show a dramatic reduction in the production of brown dwarfs and other low-mass stars \citep{Offner2009}, due to an increase in gas temperatures reducing fragmentation \citep{Krumholz+2007,Peters2010c}. More of the available gas mass ends up being accreted by the fewer, larger stars formed, and the fragmentation that does occur takes place in optically thick self-shielding discs \citep{Krumholz+2007,Peters2010a,Peters2010c}. The fact that radiation affects the mass spectrum in simulations of molecular cloud clumps has obvious implications for the shape of the initial mass function, for example the suppression of excessive brown dwarf formation \citep{Bate2009,Krumholz+2010,Peters2010c}.

Massive stars also emit prodigious amounts of UV radiation \citep{Hoare+2007,Beuther+2007} creating expanding HII regions. The hot ($10^4$ K) gas expands into the colder ($10^2$ K) surrounding low-pressure gas, creating another feedback mechanism and ionized region that may contribute to the destruction of molecular clouds \citep{Keto2002,Keto2003,Keto2007,Matzner2002,Peters2010a,Peters2010c}. HII regions can be observed by their radio continuum emission \citep{MezgerHenderson1967}, or by their recombination lines (e.g.\ \citet{WoodChurchwell1989} use the H76$\alpha$ line). More recently, observations have shown time variability in HII regions \citep{Franco-HernandezRodriguez2004,Rodriguez+2007,Galvan-Madrid+2008,gomezetal08}. \citet{Franco-HernandezRodriguez2004} have suggested that such observed time-variability may be due to the changes occurring in the source of the ionizing radiation, though it may also be due to increased absorption in the rapidly-evolving core of the nebula. \citet{Peters2010b} present a technique for using synthetic radio maps to study the time-evolution of stars forming in a cluster environment and variability in the morphology and size of HII regions. Analysis of these simulations by \citet{Peters2010a} and \citet{Galvan-Madrid+2011} confirmed variability in the flux and size measurements of HII regions, which in a few cases might be observable on timescales of $\sim 10$  years. They also noted that positive changes were more likely to occur than negative changes, i.e.\ that most of the flux variations were increases rather than decreases.

To further explore the impact of radiative feedback and the possible variability in HII regions, simulations must be equipped with good protostellar models. These have been investigated by \citet{PallaStahler1991}, \citet{PallaStahler1992}, \citet{Nakano2000}, \citet{McKeeTan2003}, \citet{Offner2009} and \citet{HosokawaOmukai2009}, among others. It is clear from these models, that the evolution of a protostar depends heavily on the mass accretion rate. Among other things, they show that the radius of the protostar may grow or contract depending on the stellar evolutionary stage. With a radius that can change significantly during the pre-main-sequence lifetime of the star, the effective temperature can also be expected to vary significantly. To study this, we simulate the formation of a cluster of stars inside a molecular cloud. We equip the stars with one of two prestellar models based on the ones described in \citet{Peters2010a} and \citet{Offner2009}, each with its own characteristics. The \citet{Offner2009} model has already been used to study star cluster formation in \citet{Krumholz+2011}, though with different initial conditions. Ours is the first simulation with the protostellar model to also include the effects of ionizing radiation and HII region formation. We connect the model to a radiative transfer method that computes the heating and ionization due to radiation from the stars formed in the simulation.

The differences between the two models is explained in \ref{sec:protmod}, but the key difference is that the \citet{Offner2009} model treats the evolution of the radius and luminosity self-consistently. The choice of stellar model affects the early evolution of stars in a cluster, and may have repercusions for the final mass spectrum. Though not entirely conclusive, we find that reduced heating and ionization in the early stages of star formation when using the \citet{Offner2009} model resulted in a more equitable mass distribution. With the \citet{Peters2010a} model, the cluster came to be dominated more by a single star about $14\Msun$ more massive than the next largest, compared to a $\sim7.5\Msun$ gap in the \citet{Offner2009} model simulations.

Other effects of the self-consistent prestellar modeling are delayed ionization of the cluster gas by 3\% of a freefall time (17.7 kyr), and delayed heating of the cluster gas by 1\% of a freefall time (5.9 kyr). 

Our numerical approach is described in Section \ref{sec:numerical_methods}. In Section \ref{sec:results} we list our results for the early evolution of star clusters with massive stars. Our assessment of the impact of protostellar modeling we discuss in Section \ref{sec:discussion} and summarize our findings in Section \ref{sec:conclusion} with a view to future simulations.

\section{Numerical Methods}\label{sec:numerical_methods}

We perform numerical simulations using the FLASH hydrodynamics code \citep{Fryxell2000} in its version 2.5. It is an adaptive-mesh refinement code that solves the gas-dynamic equations on an Eulerian grid and includes self-gravity, cooling by dust and by molecular lines \citep{Banerjee+2006}, and radiative transfer. It has been modified to include Lagrangian sink particles \citep{Banerjee2009, Federrath2010} to represent (proto)stars, and a raytracing scheme to handle ionizing and nonionizing radiation feedback from stars originally developed by \citet{Rijkhorst}, then extended and optimized by \citet{Peters2010a}. They also testing the code against handling a D-type ionization front, comparing it to the approximate solution found by \citet{Spitzer1978}, while the code's ability to handle R-type ionization fronts has already been tested by \citet{Iliev2006}. Accretion rates onto sink particles are calculated based on a single time step. FLASH does not have adaptive time steps, so every refinement level advances with the same time step.

The opacities for the non-ionizing radiation are the same as in \citet{Peters2010a}. We use Planck mean opacities as interpolated from the \citet{Pollack+2004} data by \citet{Krumholz+2007}. They assume that the radiation temperature is equal to the gas temperature because their core is optically thick. We make the same approximation using the assumption that the star will be embedded in an (unresolved) dense envelope of gas through which the stellar radiation must propagate before entering the scales of our simulation, thereby changing its spectrum accordingly.

We subsequently added an additional module to handle the protostellar evolution of our sink particles, which is based on a subgrid physics model described in detail in \citet{Offner2009}. The protostellar model connects directly to the radiation module so that stellar surface temperatures and stellar radii are handled self-consistently. The physical stages of this model are listed in Table \ref{table:protostellar_stages_paper}.

\subsection{Protostellar models}
\label{sec:protmod}

The radiative feedback model is coupled directly to the sink particles. Rays are cast outwards from each sink particle and the column density along each ray computed using the hybrid-characteristics scheme described in \citet{Rijkhorst}. At each cell in the computational domain, the photoionization rate and heating rate are calculated. These are set by the specific mean intensity along the ray,
\begin{equation}
J_\nu(r) = \left(\frac{r_{\textrm{star}}}{r}\right)^2 \frac{1}{2c^2} \frac{h\nu^3 \exp\left[-\tau_{\textrm{ion}}(r)\right]}{\exp({h\nu/k_{\textrm{B}}T_{\textrm{star}}})-1} \mathrm{ ,}
\end{equation}
which depends on a knowledge of the radius of the star $r_{\textrm{star}}$.

The stellar radius depends on the choice of stellar model. The simplest stellar model would be to assume that all stars are ZAMS stars and use a lookup table, such the one by \citet{Paxton2004}, to retrieve the radius for a star residing in a particular mass bin of the table. Such a table will also contain surface effective temperatures for ZAMS stars. The intrinsic stellar luminosity is then found from $L_{\textrm{int}} = 4 \pi R_*^2 \sigma T_{\textrm{eff}}^4$.  This model may be acceptable for most circumstances, but breaks down for when attempting to model pre-main-sequence stars. If one treats these low-mass stars as ZAMS stars, the model will underestimate their radii and overestimate their surface temperatures. It will also lead to an overestimation of the accretion luminosity $L_{\textrm{acc}} = GM\dot{M}/R$.

In \citet{Peters2010a}, a kind of ``augmented ZAMS'' model is used, which we'll refer to as A-ZAMS throughout the paper. This prestellar model uses a ZAMS description as detailed above when calculating the stellar radius and instrinsic luminosity of stars. To avoid overestimating the accretion luminosity, a separate accretion radius is calculated. This is achieved by referencing the pre-main-sequence tracks computed by \citet{HosokawaOmukai2009} for mass accretion rates between $10^{-6} \Msun$/yr and $10^{-3} \Msun$/yr and then interpolating between them based on the current mass accretion rate for the star. The advantage of this model is that it is relatively straightforward to implement and prevents grossly overestimating the accretion luminosity, which dominates the total luminosity of a star during its early lifetime. The disadvantage of this model is that it is not self-consistent and relies on two separate radii being computed or retrieved from a table. The accretion radius, found by interpolation to tracks of constant accretion rate, is sensitive to fluctuations the accretion rate. A rapidly fluctuating accretion rate means the accretion radius will fluctuate with equal rapidity---and unphysical consequence.

An alternative approach is to use the self-consistent evolving protostellar model developed by \citet{TanMcKee2004} and described in detail in \citet{Offner2009}. Stars are modeled as polytropes and every sink particle in our simulation is assigned several additional properties: a stellar radius $r_{\textrm{star}}$, an intrinsic luminosity $L_{\textrm{int}}$, a polytropic index $n$, an unburned deuterium mass $m_d$, and a nuclear burning evolutionary stage. At every timestep in our simulation, we evolve this handful of variables according to the equations given in \citet{Offner2009}. The model is based on a one-zone protostellar evolution model introduced by \citet{Nakano1995} and further developed by \citet{Nakano2000} and \citet{TanMcKee2004}.

We refer to this prestellar model as the ``evolving protostar'' model, to distinguish it from the A-ZAMS model employed in \citet{Peters2010a} and used for comparison here. It is so called because the stellar properties are co-evolved with the rest of the simulation instead of calculated on-the-fly.

When a sink particle's mass exceeds $0.1 \Msun$, we activate our protostellar evolution code and initialize the radius and polytropic index respectively as
\begin{eqnarray}
r = 2.5 \; \Rsun \left( \frac{\Delta m / \Delta t}{10^{-5} \; \Msun \; \textrm{yr}^{-1}} \right)^{0.2} \\
n = 5 - 3 \left[ 1.475 + 0.07 \; \log_{10} \left( \frac{\Delta m / \Delta t}{\Msun \; \textrm{yr}^{-1}} \right) \right]^{-1}
\end{eqnarray}
For cool stars, the Hayashi limit sets the luminosity, but above this a main sequence luminosity is assumed. Thus, $L_{\textrm{int}} = \max(L_H,L_{\textrm{ms}})$, with $L_H = 4 \pi R^2 \sigma T_H^4$ and $T_H = 3000$ K. ZAMS values for the radius and luminosity are computed using the fitting formulas by \citet{Tout1996}.

Apart from initializing our model at a higher starting mass, the only other significant difference is that we take accretion luminosity to be $L_{\textrm{acc}} = GM\dot{M}/R$. The evolving protostar model treats accretion onto the disc, with an associated luminosity of $L_{\textrm{disc}} = (1/2)\, GM\dot{M}/R$ (standard for an alpha disc), and surface accretion, with luminosity $L_{\textrm{acc}} = (1/4)\, GM\dot{M}/R$. This is due to the assumption that some of the energy is being used to drive a wind. We do not make this assumption. We also rely on tables of polytropic stellar parameters that we computed ourselves. In all other respects, our protostellar model follows the one described in \citet{Offner2009}.

\begin{figure}
\includegraphics[width=84mm,bb=0 0 504 504]{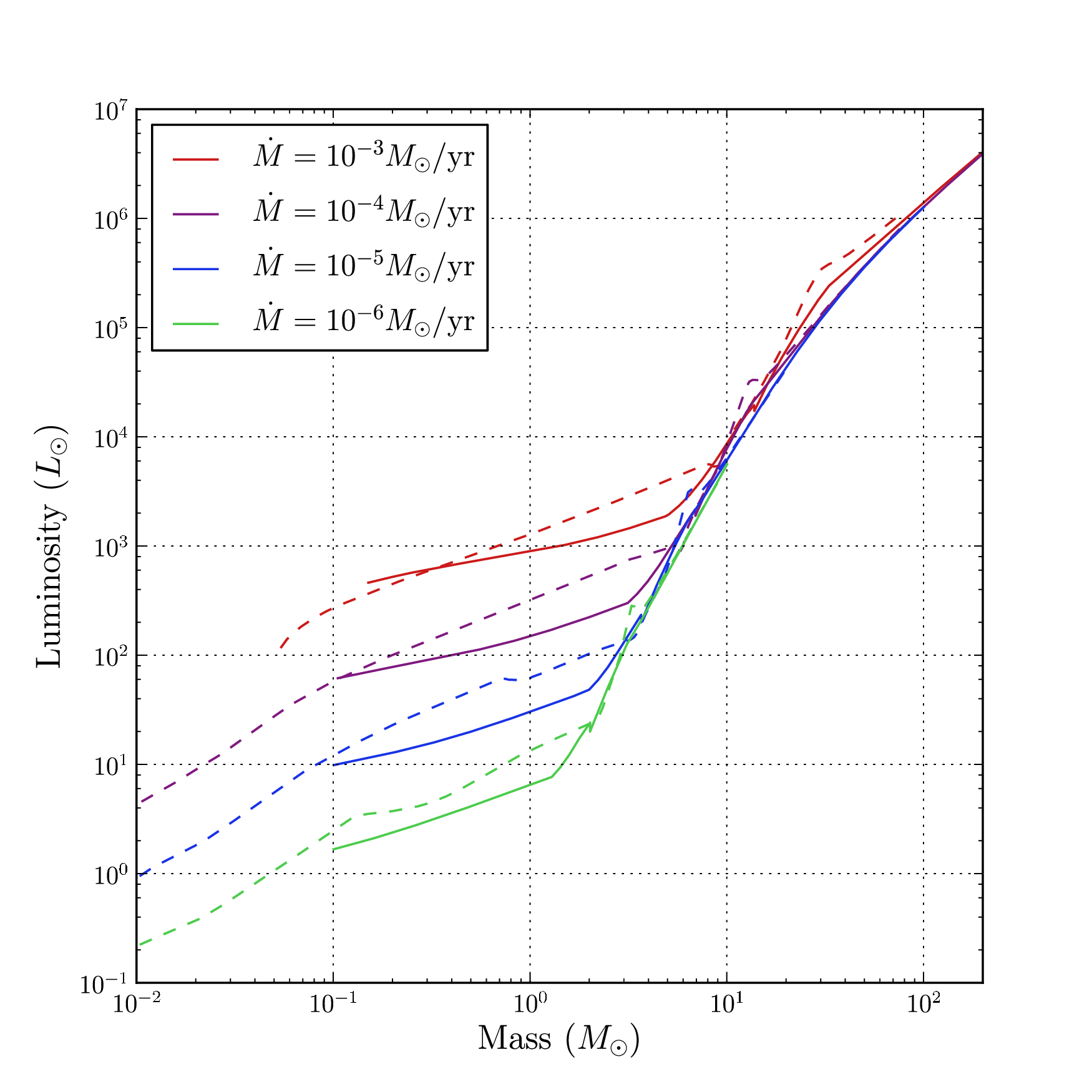}
\caption{Luminosity evolution for protostars accreting mass at various rates. The solid lines show the intrinsic (stellar) luminosity following the evolving protostar model, whereas the dashed lines show the luminosity derived from stellar structure modeling by \citet{HosokawaOmukai2009}.}
\label{fig:masslum_relation}
\end{figure}

\begin{figure}
\includegraphics[width=84mm,bb=0 0 504 504]{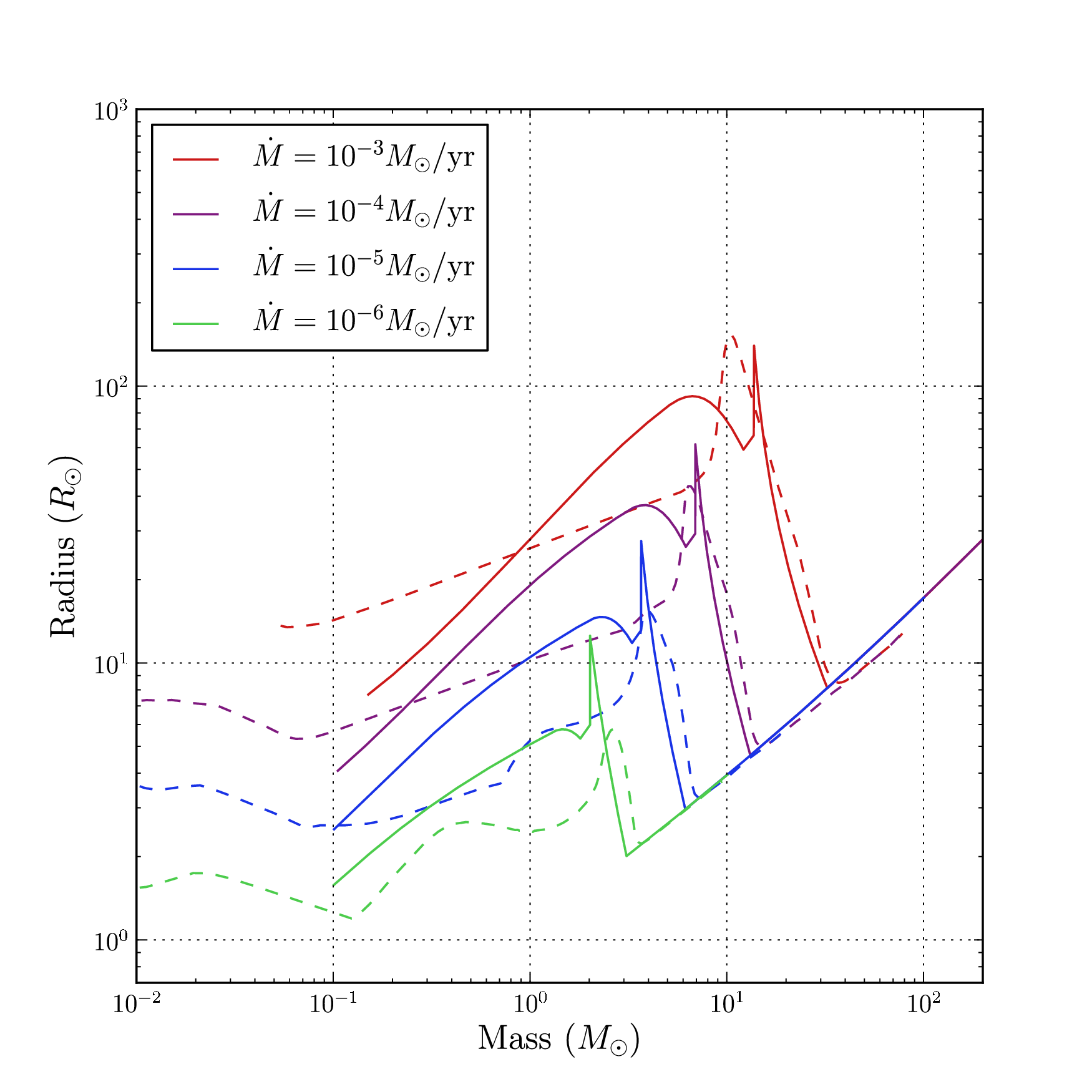}
\caption{Mass-radius relation for accreting protostars. The solid lines show the stellar radius following the evolving protostar model, whereas the dashed lines show the radius derived from stellar structure modeling by \citet{HosokawaOmukai2009}. The one-zone model approximates the stellar structure results to within a factor of $\sim2$.}
\label{fig:massradius_relation}
\end{figure}

\begin{table*}
\begin{minipage}{126mm}
\caption{Runtime parameters of the clustered star formation simulations}
\begin{tabular}{lllllll}
\hline
Run & Mass & Density profile & Temp & Rotation & Stellar Model & Feedback \\
\hline
1   & $1000\Msun$ & $r^{-3/2}$ & 30 K & $\beta = 0.05$ & ``Evolving Protostars'' & Radiative; raytracing method\\
2   & $1000\Msun$ & $r^{-3/2}$ & 30 K & $\beta = 0.05$ & ``Augmented ZAMS'' & Radiative; raytracing method\\
\hline \\
\end{tabular}
\label{table:model_table}
\end{minipage}
\end{table*}

Protostars evolve through multiple distinct nuclear stages in this code during which the radius is at times expanding (such as during the early accretion phase) and at times contracting (such as during the end stage as the protostar approaches the main sequence to be a mature star). Once our stars reach the main sequence, we assign them a radius and luminosity based on the fitting formulas of \citet{Tout1996}. We neglect any special treatment of metallicity-related effects and consider only stars of solar metallicity.

To compute the ionizing flux, we take the stellar radius and surface temperature from either a table of ZAMS values (in the case of the A-ZAMS model) or read the current, evolved values from the sink particle properties computed by the protostellar code (in the case of the evolving protostar model). The flux of ionizing photons is the computed by integrating the Planck function above the threshold frequency for hydrogen ionization. The radiative feedback code computes gas heating considering both the intrinsic and accretion luminosities.

This protostellar evolution code is a one-zone model that upgrades the current treatment of sink particles in FLASH and is a more accurate representation of pre-main-sequence stars. In Figures \ref{fig:masslum_relation} and \ref{fig:massradius_relation} we compare the results of this model with the stellar structure modeling of \citet{HosokawaOmukai2009}, which is expected to be more accurate than one-zone modeling.

We compare the behaviour of our code at different accretion rates ranging between a slow $10^{-6} \Msun$/yr to a rapid $10^{-3} \Msun$/yr. These represent the typical range of accretion rates we see in our simulations and expect of stars forming in clusters within molecular clouds. The stability of the code was tested over a range of accretion rates and timestep sizes. Although our tracks do not agree perfectly with the \citet{HosokawaOmukai2009} simulations, the agreement is to within a factor of $\sim2$.

Table \ref{table:protostellar_stages_paper}, with its accompanying figure, summarizes what is described in detail in the appendices of \citet{Offner2009}. 
\begin{figure*}
\centering
\includegraphics[width=120mm]{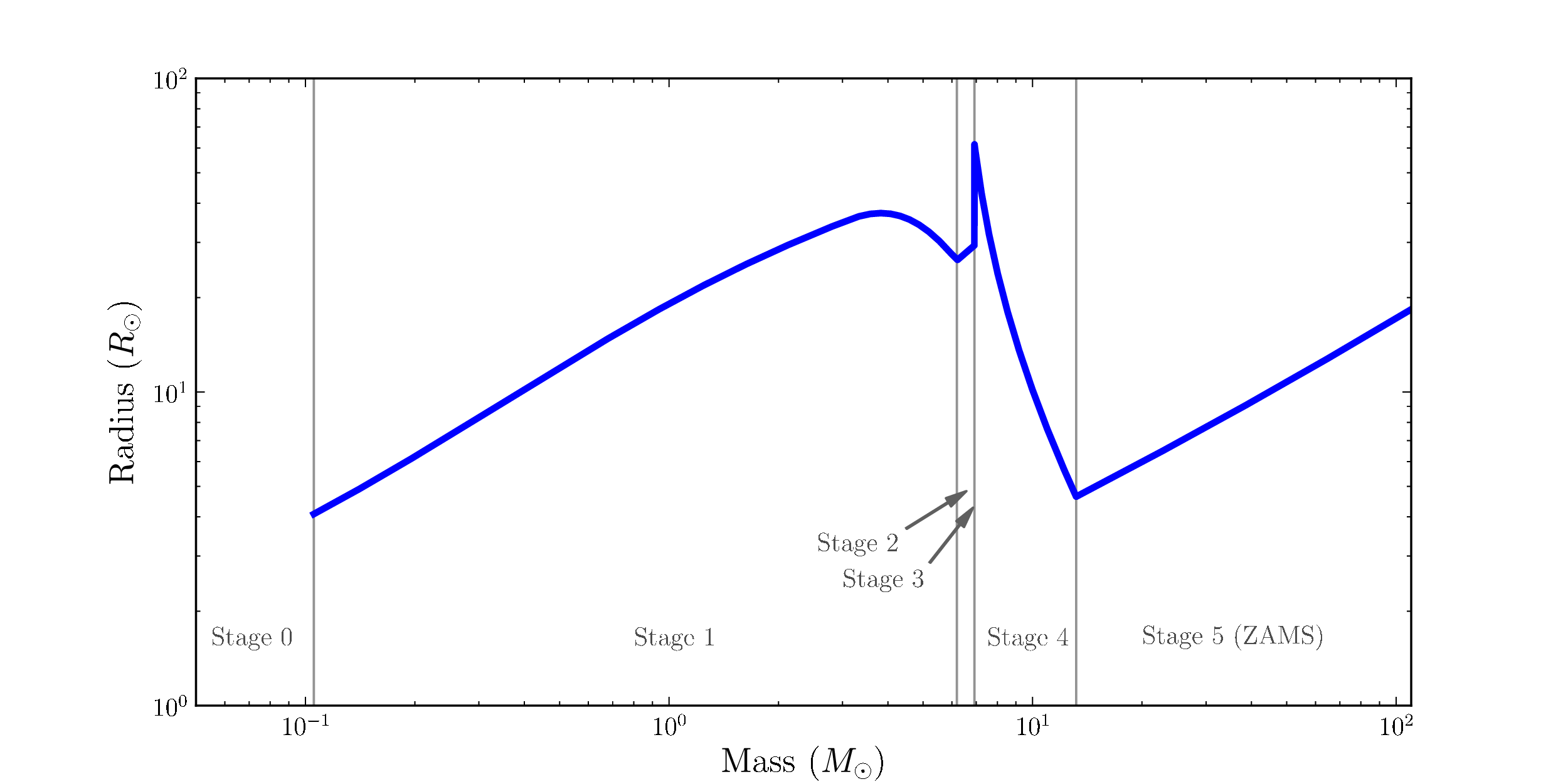}
\caption{Radius evolution of a star accreting at a steady $10^{-3} \Msun$/yr under the protostellar model of \citet{Offner2009} with stages outlined in Table \ref{table:protostellar_stages_paper}}
\label{fig:table_stages}
\end{figure*}
\begin{table*}
\begin{minipage}{126mm}
\caption{Description of the stellar evolutionary stages in the evolving protostar model}
\begin{tabular}{cll}
\hline
\multicolumn{2}{c}{Stage}		& \multicolumn{1}{c}{Features}							\\
\hline
 0 & Pre-Collapse 			& Mass $m \la 0.1 \, \Msun$ 						\\
   & 					& Cannot dissociate H$_2$ and cause second collapse to stellar densities. 	\\[1mm]
 1 & No Burning 			& Object has collapsed to stellar densities. 					\\
   & 					& $T_c$ still too cold to burn D. 						\\
   & 					& $T_c \lesssim 1.5 \times 10^6$ K 						\\
   & 					& Radiation comes purely from gravitational contraction. 			\\
   & 					& Star is imperfectly convective. 						\\[1mm]
 2 & Core D burning at fixed $T_c$ 	& Temperature reaches required $T_c \sim 10^6$ K to burn deuterium.		\\
   & 					& D burning acts as a thermostat keeping temperature constant. 			\\
   & 					& Star is fully convective. 							\\[1mm]
 3 & Core D burning at variable $T_c$ 	& D is exhausted. 								\\
   & 					& Core temperature now rising again. 						\\
   & 					& Star remains fully convective. 						\\
   & 					& Accreted D dragged down to core and burned. 					\\
   & 					& Rising core temperature reduces opacity. 					\\
   & 					& Convection in the stellar core eventually shuts down. 			\\[1mm]
 4 & Shell D burning 			& Star core changes to a radiative structure, swelling the radius.		\\
   & 					& D burns in a shell around the core. 						\\
   & 					& After initial swelling, radius contracts down to a ZAMS radius		\\[1mm]
 5 & Main Sequence 			& Star has contracted enough for $T_c$ to reach $\sim 10^7$ K 			\\
   & 					& Hydrogen ignites and star stabilizes onto the main sequence. 			\\
\hline
\end{tabular}
\label{table:protostellar_stages_paper}
\end{minipage}
\end{table*}

\subsection{Initial conditions}\label{sec:initial_conditions}

The strength of the radiative feedback code we employ lies in its ability to produce realistic HII regions. It was believed, however, that since the radius and stellar luminosity of young protostars are represented by ZAMS-equivalent values in \citet{Peters2010a}, that the ionizing flux would be overestimated. To study whether this was indeed the case, and also what impact a different prestellar model would have generally, we simulated a collapsing molecular cloud clump with each of the two prestellar models.

In the first case, we chose to repeat the cluster simulations described in \citet{Peters2010a} with similar initial conditions, but at a slightly lower resolution. Because we use the same FLASH code, sink particles, and radiative feedback code, we can isolate the effect of including a protostellar evolution model. We begin with a $1000 \Msun$ self-gravitating clump of molecular hydrogen at an initial temperature of 30 K. The cloud is in solid body rotation with a ratio of rotational to gravitational energy of $\beta = 0.05$. Our simulation box is 3.89 pc on a side. At maximal refinement, the grid size is 196 AU. The density profile features a flat central region extending out to a radius of 0.5 pc, then falling off according to an $r^{-3/2}$ power law. The central density is $\rho_c = 1.27 \times 10^{-20}$ g cm$^{-3}$. The density drops off until reaching an ambient cutoff density of $\rho_{\textrm{ext}} \approx 9.76 \times 10^{-23}$ g cm$^{-3}$. Sink particles have a radius of 1175 AU, or 6 times the grid size at maximal refinement. The cut-off density for sink particle creation is $4.4 \times 10^{-17}$ g cm$^{-3}$.

We note that clumps of this size and mass are expected to be turbulent \citep{Blitz1993,Evans1999,Williams+2000}. However, in order to build up physical understanding of the complex process of cluster formation, we follow \citet{Peters2010a} in this study and ignore turbulence so that we can isolate the important radiative feedback effects. Turbulence will be added in subsequent papers.

We show the results of two of our simulations: one using the A-ZAMS approach for stellar effective temperature and stellar radius, and a second with the evolving protostellar model.

As the simulation progresses, the original mass profile of the clump quickly disappears as it undergoes gravitational collapse to produce a rotating central disc. Stars, represented by sink particles, are allowed to form when the local conditions satisfy the criteria described in \citet{Federrath2010}.

\section{Star formation and feedback in the cluster environment}\label{sec:results}

We investigate what difference protostellar modeling makes to the overall evolution of the cluster. The most important consequence of the improved hybrid characteristics raytracing code employed by \citet{Peters2010a} is that it allows for the realistic simulation of HII regions, with ionization, heating, and shadowing effects built in. One of the most important consequences of accurate pre-main-sequence modeling is that it tempers the ionization and heating in the early stages of star formation.

\begin{figure}
\includegraphics[width=84mm,bb=0 0 540 648]{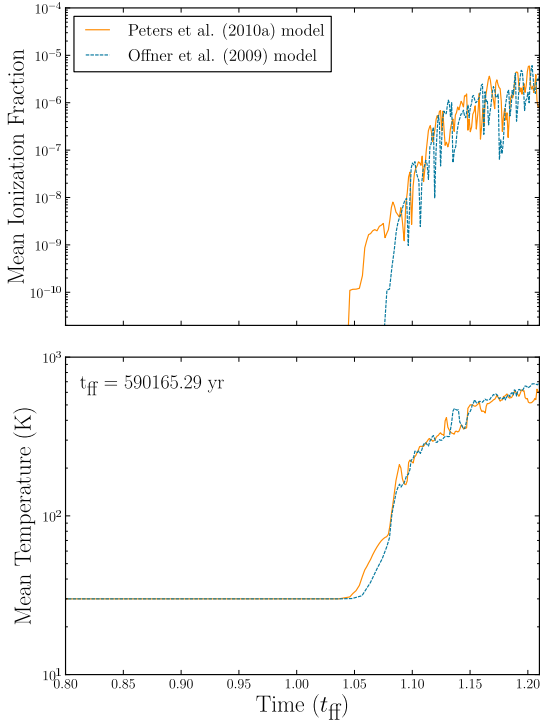}
\caption{A comparison of the mean ionization fraction and mean temperature in cluster simulations with different prestellar models. In each case, the mean is calculated by finding the volume-weighted average. Values are only meaningful in a relative sense, as the simulation volume is large (side length $\sim 3.8$pc) and the most active region is the inner cubic parsec.}
\label{fig:compare_hyd_temp_avgs}
\end{figure}

To study this effect, we look at two variables: mean ionization in our simulation box, and mean gas temperature. Figure \ref{fig:compare_hyd_temp_avgs} shows these two measurements as functions of time in our simulation. Time is measured in units of global freefall time, or $t_{\textrm{ff}} \approx 590,000$ years. With the model of \citet{Peters2010a}, sink particles follow a ZAMS model for the stellar radius and intrinsic luminosity, which means that they are hotter and more compact than true pre-main-sequence stars. This causes them to release more ionizing photons, compared to the protostellar case. The onset of ionization in this case leads the evolving protostar case by about 0.03 freefall times, or about 17.7 kyr. The onset of star formation in both simulations occurs at around 1 freefall time. For both cases, after 1.1 freefall times, the largest star in either simulation is at $\sim20 \Msun$ and dominates the UV output of the cluster, resulting in comparable mean ionization for both cases.

When we consider mean temperature instead of mean ionization, the leading effect by the ZAMS-based model is still there, only less pronounced. Major heating of the gas in this case leads the evolving protostar model case by close to 0.01 freefall times, or about 5.9 kyr. The first star to form in a cluster tends to grow to be among the largest stars in the cluster and dominate the heating and ionization. This suggests that accurate protostellar modeling is most important in the early stages of a cluster simulation, and for low-mass stars. \citet{Offner2009} showed that radiation even from low-mass stars has a significant effect on the gas heating and formation of brown dwarfs.

\begin{figure*}
\includegraphics[width=180mm]{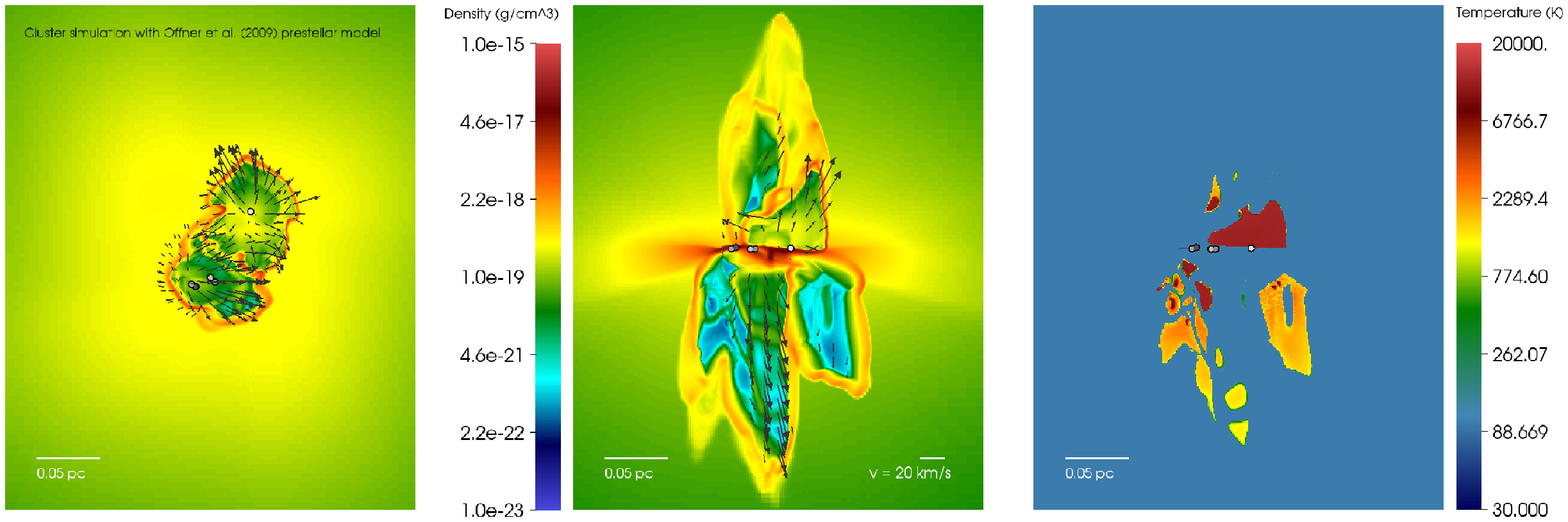}
\includegraphics[width=180mm]{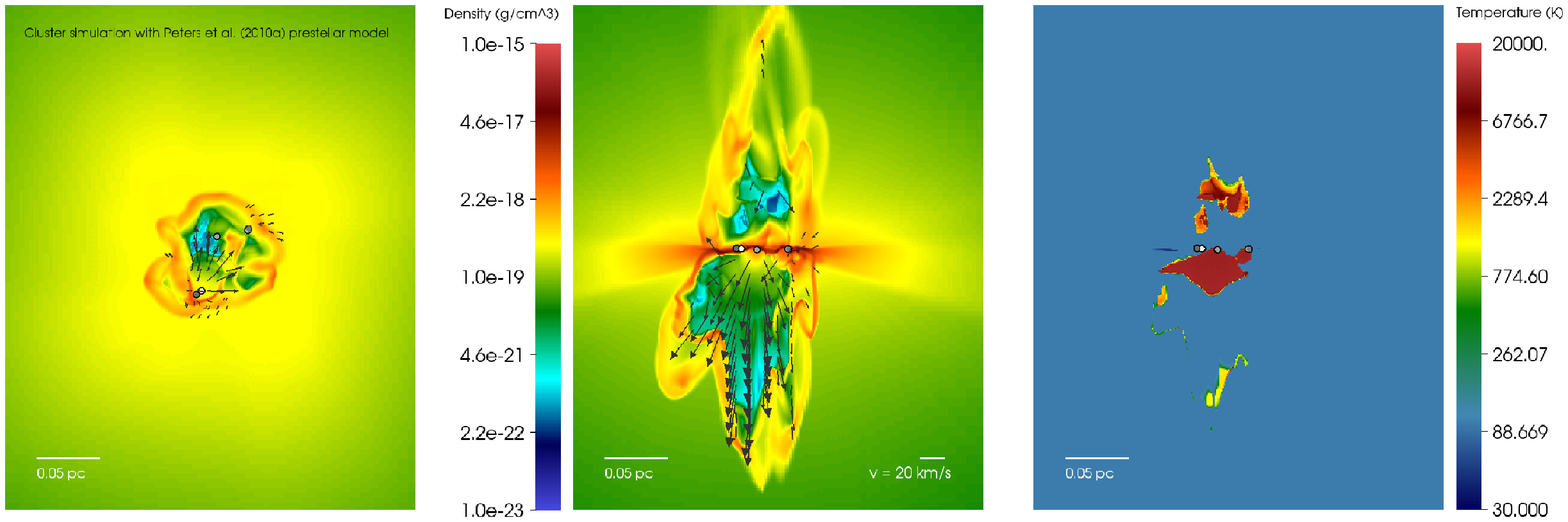}
\caption{The two rows show the results from the two prestellar models tested, with the A-ZAMS model of \citet{Peters2010a} in the top row and the evolving protostar model of \citet{Offner2009} in the bottom row. Each is shown near the end of the simulation, after about 1.21 freefall times (714 kyr). In each row, the panels show, from left to right: the gas density in a horizontal slice through the midplane, the gas density in a vertical slice in the centre of the simulation box, and the gas temperature in the same slice. Scale bars indicate the physical sizes and the speeds represented by vectors in the gas density panels. The scale for these vectors is the same for both side views and top-down views. Stars are indicated by black circles.}
\label{fig:cluster_views_dens+temp}
\end{figure*}

To get a visual sense of the gas dynamics and configuration of the cluster, we visualize the gas density by taking slices through our simulation box. Zoomed-in views of the cluster are shown in Figure \ref{fig:cluster_views_dens+temp}. The simulation box is actually about 3.8 pc across. Here we show the central region, at about 0.5 pc across. The upper row in the figure shows the simulation results with the evolving protostar model, while the lower row shows the A-ZAMS results. In each row, the panels show: gas density in a horizontal slice through midplane of the simulation box (left), gas density in a vertical slice showing the cluster edge-on (centre), and gas temperature in the same vertical slice (right). Gas temperature is discussed in Section \ref{sec:ionization_and_temperature}. The two density panels also show velocity vectors for the high-velocity gas. The fastest-moving gas travels at close to 30 km/s. Gas densities range from $10^{-23}$ to $10^{-15}$ g/cm$^3$. The hollowed-out HII regions, where the gas is largely ionized, expand outwards above and below the disc as a kind of fountain before falling back onto the disc.

Sink particles indicating the locations of stars are marked with black-rimmed gray points. The side view shows the stars to be confined to the disc while the top-down shows the stars packed in a tight cluster. The separation between stars nowhere exceeds 0.1 pc. During the course of the simulation, stars are seen to be dynamically interacting, exchanging angular momentum, forming and breaking apart binaries.

These snapshots of the simulation are taken at around 1.21 freefall times in each case, near the end of the simulation. At this stage, about 714 kyr have elapsed since the beginning of the simulation, with the onset of star formation having occurred at around 600 kyr. At this stage, both model results look similar in many ways: the stars are in a densely-packed cluster, and each cluster has produced an expanding HII region. The HII regions in each figure are approximately the same size, although amorphous and variable. They do not seem to be affected by our choice of prestellar model. This is because of how each cluster has become dominated by massive stars already evolved onto the main sequence, and the differences between prestellar models has vanished. The stars are all releasing copious amounts of ionizing radiation, driving the evolution of these HII regions.

\subsection{Binaries}

\citet{ZinneckerYorke2007} state that massive stars occur more frequently in binaries relative to low-mass stars. Lacking turbulence and magnetic fields, our molecular gas clumps do not represent the true initial conditions for cluster formation, but the stars in our simulations to form binaries. There is no reason to suspect that the choice of protostellar model has any effect on binary formation or binary mass ratios. Lacking ensemble averages, we cannot make any special claims, but report that of the 5 stars formed in each of our simulation, 4 stars end up in binaries. Dynamical encounters between stars cause binaries to form, break apart, and reform. The final  mass ratios of the two pairs each simulation were 3.74 and 1.61 with the Offner model. The Peters model simulation saw mass ratios of 3.38 and 1.92. The final masses of the stars formed in each simulation are reported in Table \ref{table:final_masses}.

\subsection{Accretion Histories}

\begin{figure}
\includegraphics[width=84mm,bb=0 0 576 720]{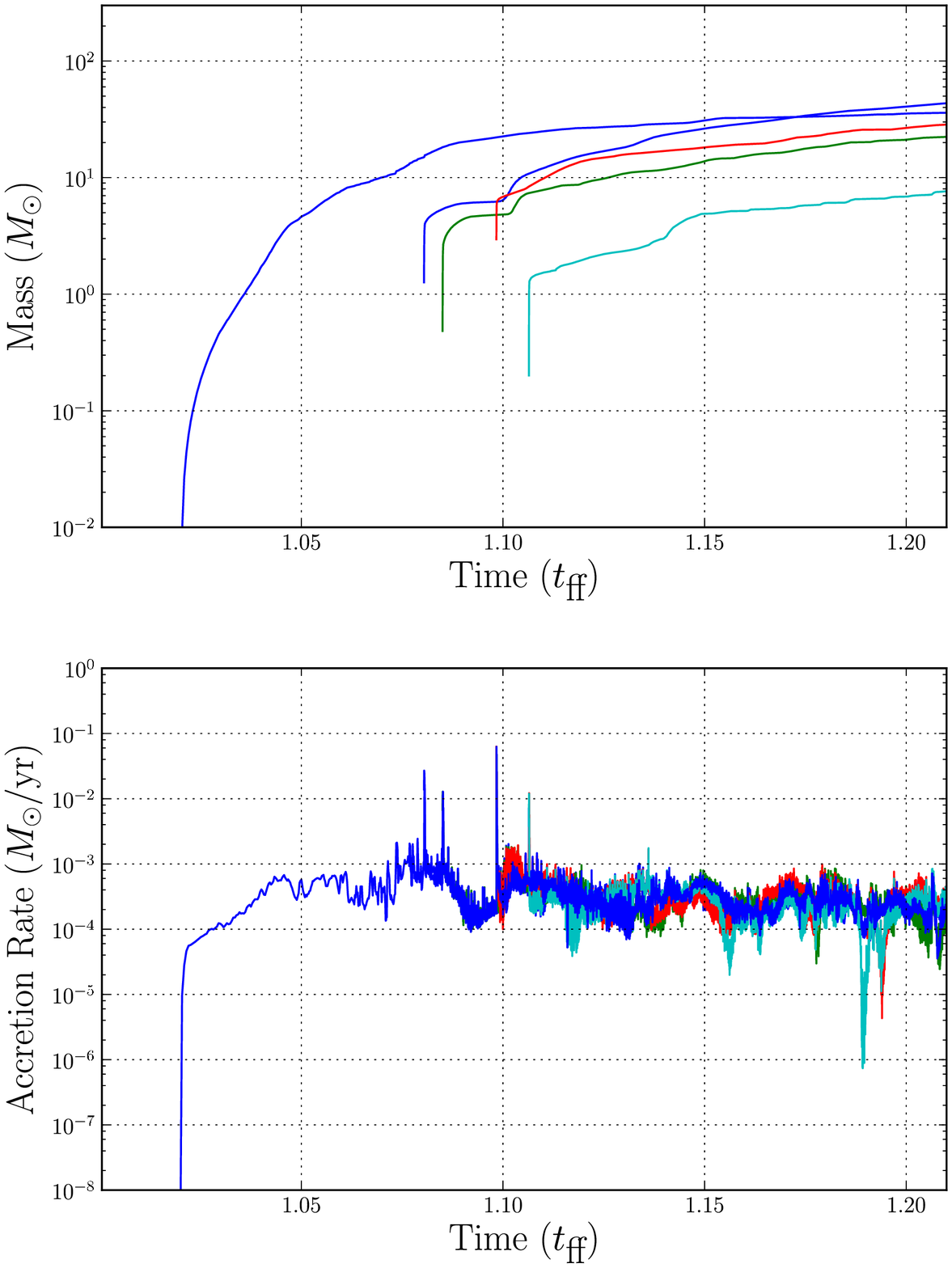}
\caption{Accretion histories of stars formed in the cluster simulation of the evolving protostars setup. The upper panel shows the mass of each particle as a function of time. The lower panel shows the accretion rate in units of $\Msun$ yr$^{-1}$ as a function of time. The dynamical time is about 0.59 Myr.}
\label{fig:protostellar_accretion_history}
\end{figure}
\begin{figure}
\includegraphics[width=84mm,bb=0 0 576 720]{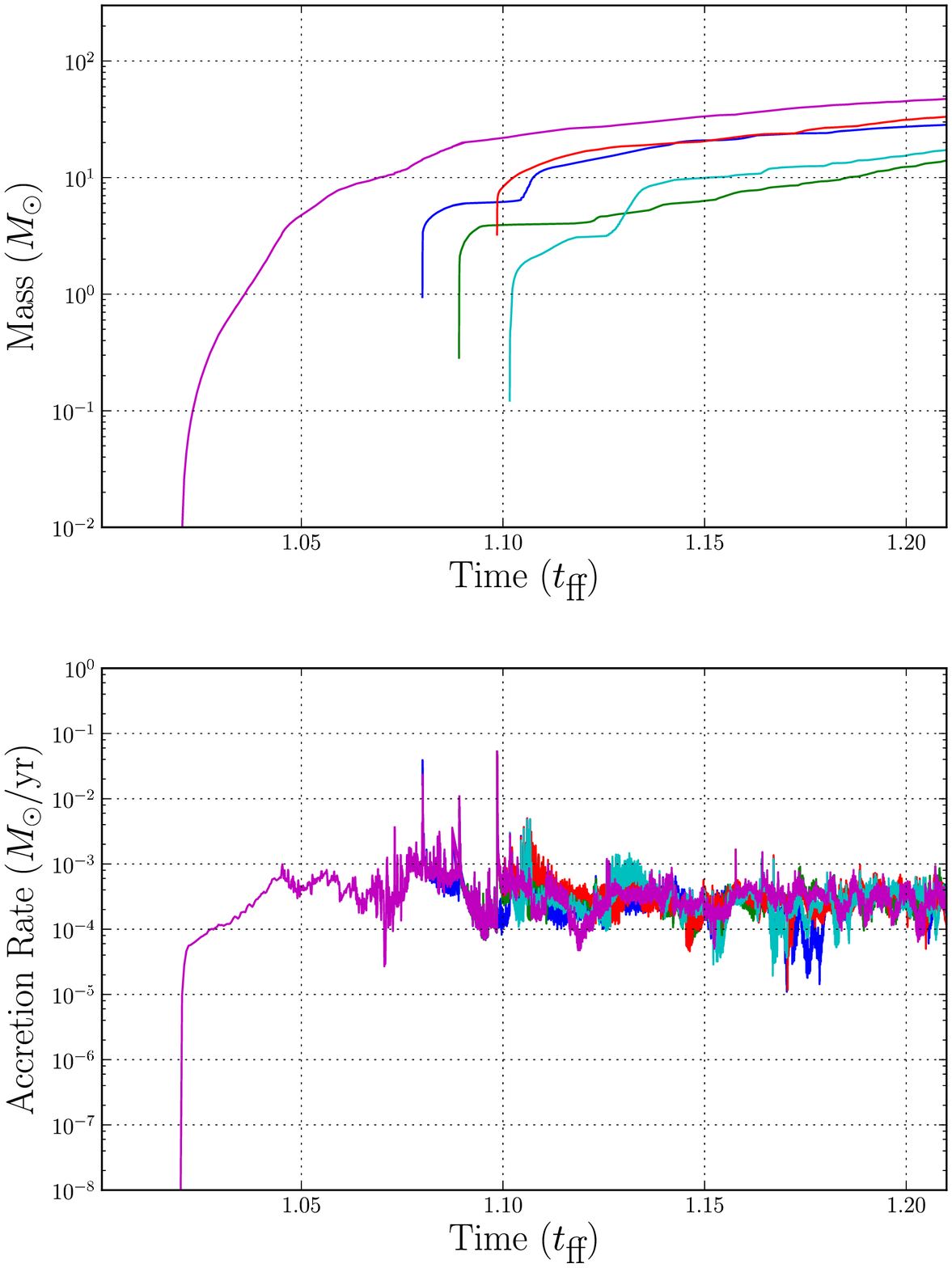}
\caption{Accretion histories of stars formed in the cluster simulation of the A-ZAMS setup. The upper panel shows the mass of each particle as a function of time. The lower panel shows the accretion rate in units of $\Msun$ yr$^{-1}$ as a function fo time. The dynamical time is about 0.59 Myr.}
\label{fig:noproto_accretion_history}
\end{figure}

We now compare the simulations with a focus on the accretion histories of the sink particles. \citet{Peters2010a} have shown that the gas surrounding the centre of the cluster would fragment and result in a highly variable accretion rate. We see this in Figures \ref{fig:protostellar_accretion_history} and \ref{fig:noproto_accretion_history}, where we show in the two panels the accretion histories of every sink particle formed in our simulation along with their accretion rates.

In the lower panel we see the accretion rate of each star, and for most of the stars in our simulation, the accretion rate remains between $10^{-4}$ and $10^{-3} \Msun$/yr.

The upper panel in Figure \ref{fig:protostellar_accretion_history} shows the growing masses of each of the stars in our protostellar model simulation. Star formation does not really commence until after the first dynamical time (freefall time)---about 0.59 Myr for our simulation setup. There seems to be a burst of star formation after $t \approx 1.10 t_{\textrm{ff}}$. Interestingly, the most massive star is not the first star in our simulation, but it is overtaken in mass by the second star, which reaches a final mass of about $43.5 \Msun$. The others reach final masses of approximately 36.0, 28.5, 22.3, and 7.6 $\Msun$. The average mass of these 5 stars is 27.6 $\Msun$. We were able to run the evolving protostars simulation longer than the A-ZAMS case. During this extra time, 3 additional stars formed and accreted about $1 \Msun$ of material each, but we do not use this additional data in our comparison with the A-ZAMS case.

\begin{table}
\caption{Final stellar masses after 1.21 freefall times (714 kyr) in each cluster simulation, in units of solar masses, comparing the different prestellar model results.}
\begin{center}
\begin{tabular}{cc}
\hline
Evolving Protostars & Augmented ZAMS\\
\hline
43.5 & 47.3\\
36.0 & 33.3\\
28.5 & 28.4\\
22.3 & 17.3\\
 7.6 & 14.0\\
\hline
\end{tabular}
\end{center}
\label{table:final_masses}
\end{table}

In the A-ZAMS case, shown in Figure \ref{fig:noproto_accretion_history}, the final masses of the stars are 47.3, 33.3, 28.4, 17.3, and 14.0 $\Msun$. The average mass of these 5 stars is 28.1 $\Msun$. It is difficult to draw firm conclusions about the impact of a protostellar model on the population dynamics of a cluster. We would need to complete longer simulations under more realistic initial conditions (including turbulence). The evolving protostars run experienced a second wave of star formation, but when we restrict ourselves to comparing only the first 1.21 $t_{\textrm{ff}}$ in each case, we find that they have almost the same average mass.

Interestingly, though, the evolving protostars case had 4 stars with masses greater than $10 \Msun$. These were all more closely packed (smaller variance), than the four most massive stars in the A-ZAMS case. We propose that the reduced initial heating and ionization from the self-consistently evolved pre-main-sequence stars results in a more equitable partition of mass between the massive stars. The most massive star in this simulation outranks the second largest by about $7.5 \Msun$. By comparison, the leading star in the A-ZAMS simulation exceeded the next most massive star by $14 \Msun$, or nearly double. Further simulations with different initial conditions are required to confirm whether this choice of prestellar model will always have such an impact.

These are the results of only a single simulation in each case, so it is difficult to say that this difference in massive spectrum is highly significant, especially given that the average mass of each cluster is similar and our simulations did not contain turbulence. By other measures, such as the average ionization, mean temperature, or HII region morphology, the two prestellar models converged and gave similar results. The mass spectrum shows a similar average mass of $\sim28 \Msun$, but with the evolving protostar model having both a smaller peak mass and smaller difference in mass between the two top stars relative to the A-ZAMS model.

We were able to run the protostellar simulation a little longer than the A-ZAMS simulation; there occurred a second burst of star formation that only appeared very late in the simulation. These stars grew to be 1.9, 1.3 and 1.0$\Msun$. The A-ZAMS run may have formed more stars if run for longer. We have run each setup for approximately two weeks on 64 processors, or approximately 21,500 CPU-hours. The protostellar simulation progressed further than the ZAMS simulation. In either case, memory or eventual code stability limited the length of the runs.

\subsection{Mass-radius relation}

\begin{figure}
\includegraphics[width=84mm,bb=0 0 576 360]{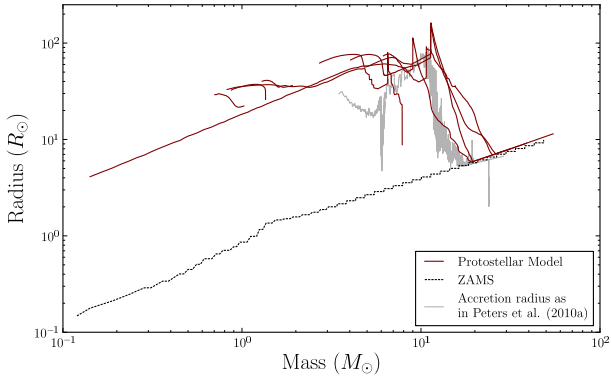}
\caption{The mass-radius relation for the stars in both cluster simulations. The black dashed line marks the stellar radius track of a sink particles in the A-ZAMS simulation. The radius is based on tabulated values of luminosities and temperatures for ZAMS stars. The gray line indicates the separately-calculated accretion radius for a single star. Red lines mark the tracks of sink particles following the evolving protostar model. This protostellar radius is used as both the stellar radius and the accretion radius.}
\label{fig:cluster_mass_radius_relations}
\end{figure}

The mass-radius relation for a star is a means of comparing different protostellar models. It is also a way of seeing the evolution of the stars in our simulation. As stars accrete mass or undergo nuclear-structural changes in their interiors, the radius reacts either by expanding or contracting. We see the evolution of the stars in our simulation represented in Figure \ref{fig:cluster_mass_radius_relations}. In this figure we compare the radii of stars from our different cluster simulations. 

We show the accretion radius for a single star in the A-ZAMS run by the gray line in Figure \ref{fig:cluster_mass_radius_relations}. The red lines in this figure are the stars following the protostellar evolution model that we have described in Section~\ref{sec:protmod}. These stars have their radius continuously evolved according to their burning state and the accretion of new material. The radius, therefore, does not fluctuate with unrealistic rapidity. Because the model has a self-consistent description of the radius, we use the same quantity to describe the stellar radius and the accretion radius, rather than computing each by different means. Protostars have radii an order of magnitude larger than a zero-age main-sequence star of equal mass. Hence, their effective temperatures and flux of ionizing photons are going to be much less (for a $1\Msun$ star, 3000K vs.\ 5000K in effective temperature, $10^{29}$ s$^{-1}$ vs $10^{39}$ s$^{-1}$ in ionizing photons). Star in simulations without protostellar modeling may excessively heat or ionize the gas during the early phases of star formation.

\begin{figure}
\includegraphics[width=84mm,bb=0 0 576 360]{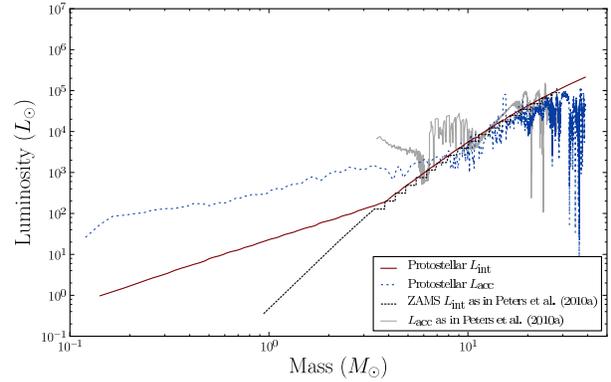}
\caption{The mass-luminosity relation for a representative star in each of the two cluster simulations. For each star, its accretion luminosity and intrinsic stellar luminosity are plotted. The red and dashed blue lines show the instrinsic stellar luminosity and the accretion luminosity, respectively, of a star in the evolving protostars simulation. The black dashed line, with its stepped appearance, represents the intrinsic luminosity of a ZAMS star, retrieved from a table of ZAMS values. Finally, the grey line shows the accretion luminosity of a star in our in the A-ZAMS simulation. The luminosity is calculated as in \citet{Peters2010a} by an interpolation of the radius to models by \citet{HosokawaOmukai2009}.}
\label{fig:cluster_mass_luminosity_relations}
\end{figure}

The evolution of the stars in each simulation is also revealed by the mass-luminosity relation, shown in Figure \ref{fig:cluster_mass_luminosity_relations}. Black and grey lines belong to a representative star in the the A-ZAMS model simulation, dark red and blue to a representative star in the protostellar simulation. Accretion luminosity, calculated as $L_{\textrm{acc}} = G M \dot{M} / R_{\textrm{acc}}$, is especially sensitive to the accretion rate $\dot{M}$ and the accretion radius $R_{\textrm{acc}}$. Stars show accretion luminosities that are up to an order of magnitude larger than the stars in the evolving protostar model simulation on account of the difference in stellar radius. Only for stars larger than about $20 \Msun$ do the differences between the two models disappear. The black jagged line indicates the main sequence luminosities from a precomputed table, which tends to underestimate stellar luminosities for protostars less than about $3 \Msun$. Protostellar luminosity of one star is given by the dark red line and protostellar accretion luminosity by the blue line. Much of the rapid fluctuation in the accretion luminosities of both simulations stems from the highly variable mass accretion rate (see Figures \ref{fig:protostellar_accretion_history} and \ref{fig:noproto_accretion_history}).

\subsection{Ionization and Temperature}\label{sec:ionization_and_temperature}

\begin{figure*}
\includegraphics[width=126mm]{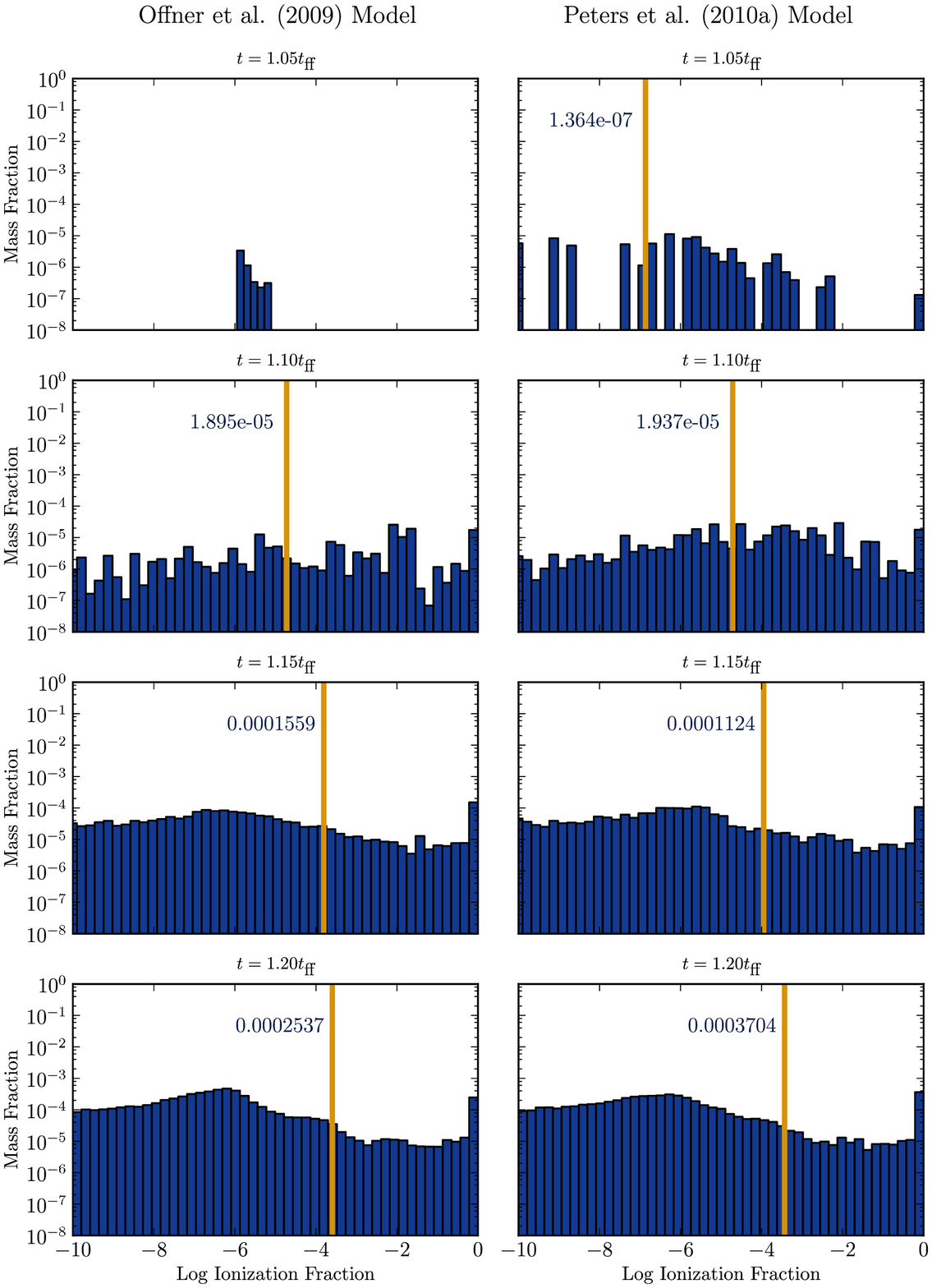}
\caption{The evolving mass-weighted ionization fraction spectrum. Compared are cluster simulations with stars running on the evolving protostar model of \citet{Offner2009} on the left and the A-ZAMS model of \citet{Peters2010a} on the right. A distribution of the total mass in the simulation box (about 1000 $\Msun$) is shown for $t = 1.05, 1.10, 1.15, 1.20 \, t_{\textrm{ff}}$. One freefall time is approximately 0.59 Myr. The yellow line indicates the mass-weighted average ionization fraction, the numerical value of which is printed to the left of the line.}
\label{fig:cluster_hyd1_distribution}
\end{figure*}

Protostars have large radii about an order of magnitude larger than equivalent-mass main-sequence stars. They may be just as luminous, and they certainly have high accretion luminosities, but it is the effective temperatures of their surfaces that determine how great the flux of ionizing photons will be, if the star emits any at all. The single greatest difference we see when simulating the evolution of a star cluster with self-consistent protostellar modeling is that when the first stars begin to form after about a dynamical time, the average gas temperature and average ionization of the gas is considerably less in the simulation involving our protostellar model (Figure \ref{fig:compare_hyd_temp_avgs}).

Figure \ref{fig:cluster_hyd1_distribution} shows the mass-weighted spectrum of the ionization fraction in both cases, with the evolving protostar model on the left and the A-ZAMS model on the right. Values for ionization fraction range from $10^{-10}$ to approximately 1 (completely ionized). Ionization fractions $\ll 1$ should not be taken too seriously, as our model includes only stellar ionizing radiation from the stars in our cluster. The figure shows the spectrum for all the gas involved in the simulation---approximately 1000 $\Msun$ in total. The thick yellow line indicates the mass-weighted average value for the ionization fraction with the value printed beside the line. Individual snapshots in time are: $t = 1.05, 1.10, 1.15, 1.20 \, t_{\textrm{ff}}$.

It is important to show how the averages change over time in Figure \ref{fig:compare_hyd_temp_avgs} because of how the mean tends to fluctuate yet the two models have similar values for all but the earliest phases of star formation. The early phase is shown in the first row, at $t = 1.05 t_{\textrm{ff}}$. Here there is a significant difference in the mean ionization fractions of the two models. The low-mass ZAMS stars are hotter and have smaller radii. There is greater early ionization seen in this case. At later stages, the distributions appear more similar as the conspicuous effects of the model disappear.

The temperature structure of the gas surrounding the cluster is shown in right panels of Figure \ref{fig:cluster_views_dens+temp}. These reveal some interesting features. The gas in the vicinity of the cluster is approximately 100 K, heating to this temperature by the nonionizing radiation coming from the cluster. We also see pockets of very hot ($10^4$ K), ionized gas in the expanding HII regions. When we study the evolution of these regions in time, we see that these pockets of high-temperature gas are very transient: forming, expanding, breaking apart, and cooling very rapidly. They are due to photoionization and photoionization heating caused by the massive stars in the cluster. \citet{Peters2010a} have attributed this flickering to the chaotic gas motions in the cluster. Gas moves toward the interior of the cluster through the disc and interacts with the ionizing radiation giving rise to many unstable morphologies that expand outwards above and below the disc. This has the appearance of flickering on relatively short timescales: less than 560 years---the temporal resolution of our simulations. Synthetic observations of the original \citet{Peters2010a} results analyzed by \citet{Galvan-Madrid+2011} have revealed this flickering visible in radio-continuum emission and have demonstrated that it is in agreement with available observations.

\section{Discussion}\label{sec:discussion}

Considering that the radii and luminosities of true protostars are vastly different from their ZAMS counterparts (i.e.~Figures \ref{fig:cluster_mass_radius_relations} and \ref{fig:cluster_mass_luminosity_relations}), it may seem surprising that our two simulations actually look so alike. For instance, the two simulations form an equal number of stars, their average mass is approximately the same, and morphology of the clump with its outflows and HII regions appear qualitatively similar in both cases. It is important to note what we are comparing. The model that we are comparing against \citep{Peters2010a} treated the stellar radius and the accretion radius separately, meaning that gas heating has two components: one due stellar radiation, and one due to the accretion luminosity. The mass-luminosity relation of Figure \ref{fig:cluster_mass_luminosity_relations} shows that the ZAMS stellar luminosity underestimates the true protostellar luminosity for pre-main-sequence stars. It also shows that the accretion luminosity, calculated as in \citet{Peters2010a} by an interpolation to the \citet{HosokawaOmukai2009} models, overestimates the true protostellar accretion luminosity. So there are two competing differences and these effects will partially cancel each other out. The result is that our evolving protostars simulation looks similar in many ways to the results of \citet{Peters2010a}. If \citet{Peters2010a} had not boostrapped the separate treatment of accretion radius on to the ZAMS model, there may have been a gross overestimation of the accretion luminosity---which dominates the total luminosity of a star during its early lifetime. The errors resulting from this overestimation could be substantial.

In our radiative feedback technique, we treat ionization separately from heating, and ionization depends solely on the effective temperatures of our stars. Since protostars have cooler surface temperatures than ZAMS stars of equal mass, there is much less early ionization. Since it is the ZAMS stars of high mass that dominate the radiation output of a cluster, the differences between our model and the ZAMS model dissapear after the early stages of stellar evolution (Figure \ref{fig:cluster_hyd1_distribution}). A side-by-side comparison the ionizing flux from stars with different stellar models will be included in a forthcoming paper.

Our simulations have a number of limitations that should be noted. They neglect the effects of radiation pressure. On large scales, radiation pressure from stellar clusters could drive galactic winds \citep{Murray+2011}. However, within our low-density $1000\Msun$ cluster, radiation pressure below the Eddington limit should not be dynamically significant \citep{YorkeSonnhalter2002, KrumholzMatzner2009}. After the first absorption/reemission event, the radiation will have been converted to infrared radiation to which the molecular cloud is largely transparent. The first absorption event is unlikely to impart a significant amount of momentum.

In our simulations, we have treated gas that was initially cold and in solid body rotation, but without any turbulence. Cluster-forming clumps in molecular clouds are observed to have supersonic turbulence, and a more realistic set of initial conditions would include turbulence. However, this might have obscured the effects of our protostellar model that we were seeking to measure. We are currently preparing to run simulations that include realistic turbulent initial conditions as well magnetic fields, which were also left out of this simulation (see, however, \citet{Peters2011} for the effects of magnetic fields on our non-turbulent initial conditions).

The protostellar model we have added to our simulations improves on previous work by adjusting the ionizing luminosity so that it matches the stellar surface effective temperatures for accreting protostars, which initially have radii larger than equal-mass stars on the main sequence. We note, however, that a full-spectrum treatment of the radiation still faces technical and computational limits that make the problem extremely challenging. As a compromise, we break the radiation into its ionizing and nonionizing components.

\section{Conclusions}\label{sec:conclusion}

Stars begin to affect their birth environments as soon as they are born through radiative feedback. We have considered the impact that pre-main-sequence modeling can have on a star cluster by comparing two different prestellar models already described in the literature. We did this by repeating the simulation of \citet{Peters2010a}. We then upgraded the FLASH code to include a protostellar evolution module based on the one described in the appendices of \citet{Offner2009}.

Each model works by equipping the stars in the simulation (``sink particles'') with a stellar radius and luminosity. The greatest difference between the two models was self-consistency. The \citet{Peters2010a} model calculated approximate stellar parameters on-the-fly, while the evolving protostar model evolved the stellar parameters self-consistently through the simulation as the stars grew and accreted mass.

In terms of the overall gas structure, HII regions, temperature structure, mean ionization fraction, or stellar binarity, the two models produced qualitatively the same results. This is because a cluster comes to be dominated by its most massive stars, which are evolved, main-sequence, highly luminous stars, regardless of the choice of stellar model. These one or two massive stars control the overall dynamics.

The differences exist in the early phase of star formation. Major ionization of the gas in the evolving protostar model lagged the \citet{Peters2010a} model by about 3\% of a freefall time, or about 17.7 kyr. Major heating of the gas lagged by about 1\% of a freefall time, or about 5.9 kyr. The difference in heating and ionization was due to the fact in \citet{Peters2010a}, the stellar radius was underestimated (a ZAMS-equivalent value was taken), when protostars have radii an order of magnitude larger than a zero-age main-sequence star of equal mass. The correspondingly higher surface temperatures resulted in excess heating and ionization in this model. When both models had stars converging onto the main sequence, the differences between the two models diminished.

It is possible that these initial differences could have had a lasting effect on the stellar population. The most massive star at the end of each simulation was 43.5$\Msun$ in the evolving protostar model, and 47.3$\Msun$ in the \citet{Peters2010a} case---a difference of 8\%. The differences in mass between the most massive star and the next largest star was 7.5$\Msun$ in the evolving protostar case and 14$\Msun$ in the \citet{Peters2010a} case. It would require further simulations, varying the initial conditions, to confirm that this is always the case.

The cluster of stars is embedded in a rotating disc of gas approximately 0.2 pc in size. The expanding HII regions above and below the disc are rapidly changing in shape and size on timescales shorter than 570 years. The physical size of these HII regions in our simulation is at most about 0.2 pc. This flickering is observed regardless of the prestellar model used.

Future simulations will have initial conditions including turbulence to model molecular clouds as realistically as possible. The stars will no longer be forming within a global disc, but rather along sheets and filaments in diverse parts of the cloud. With star formation thus spread out more in space and time, we expect the influence of individual young stellar objects on their environments to be more significant than when all stars form in a central cluster. It will be important to have the radiative feedback accurately modeled in these cases.

\section*{Acknowledgments}

We thank Mark Krumholz for clarifying some of the more technical aspects of the protostellar evolution method employed in the ORION code and described in \citet{Offner2009}. This was very helpful in developing the FLASH module following the same method. We thank Takashi Hosokawa for sharing data from \citet{HosokawaOmukai2009} with us. We also thank our anonymous referee for a very careful review of our paper that helped to significantly clarify the presentation of our results. M.K.~acknowledges financial support from the Ontario Graduate Scholarship Program. R.E.P.~is supported by a Discovery Grant from the Natural Sciences and Engineering Research Council (NSERC) of Canada. T.P.~acknowledges financial support as a Fellow of the Baden-W\"{u}rttemberg Stiftung funded by their program International Collaboration II (grant P-LS-SPII/18) and through SNF grant 200020\_137896. The FLASH code was in part developed by the DOE-supported Alliances Center for Astrophysical Thermonuclear Flashes (ASCI) at the University of Chicago. This work was made possible by the facilities of the Shared Hierarchical Academic Research Computing Network (SHARCNET: www.sharcnet.ca) and Compute/Calcul Canada.

\bibliography{simulating_protostellar}

\begin{thebibliography}{48}
\expandafter\ifx\csname natexlab\endcsname\relax\def\natexlab#1{#1}\fi

\bibitem[{{Banerjee} {et~al}\mbox{.}(2006){Banerjee}, {Pudritz}, \&
  {Anderson}}]{Banerjee+2006}
{Banerjee} R., {Pudritz} R.~E., {Anderson} D.~W., 2006, \mnras, 373, 1091

\bibitem[{{Banerjee} {et~al}\mbox{.}(2009){Banerjee}, {V{\'a}zquez-Semadeni},
  {Hennebelle}, \& {Klessen}}]{Banerjee2009}
{Banerjee} R., {V{\'a}zquez-Semadeni} E., {Hennebelle} P., {Klessen} R.~S.,
  2009, \mnras, 398, 1082

\bibitem[{{Bate}(2009)}]{Bate2009}
{Bate} M.~R., 2009, \mnras, 392, 1363

\bibitem[{{Beuther} {et~al}\mbox{.}(2007){Beuther}, {Churchwell}, {McKee}, \&
  {Tan}}]{Beuther+2007}
{Beuther} H., {Churchwell} E.~B., {McKee} C.~F., {Tan} J.~C., 2007, Protostars
  and Planets V, 165

\bibitem[{{Blitz}(1993)}]{Blitz1993}
{Blitz} L., 1993, in Protostars and Planets III, {E.~H.~Levy \& J.~I.~Lunine},
  ed., pp. 125--161

\bibitem[{{Clarke} {et~al}\mbox{.}(2000){Clarke}, {Bonnell}, \&
  {Hillenbrand}}]{Clarke+2000}
{Clarke} C.~J., {Bonnell} I.~A., {Hillenbrand} L.~A., 2000, Protostars and
  Planets IV, 151

\bibitem[{{Evans}(1999)}]{Evans1999}
{Evans}, II N.~J., 1999, \araa, 37, 311

\bibitem[{{Federrath} {et~al}\mbox{.}(2010){Federrath}, {Banerjee}, {Clark}, \&
  {Klessen}}]{Federrath2010}
{Federrath} C., {Banerjee} R., {Clark} P.~C., {Klessen} R.~S., 2010, \apj, 713,
  269

\bibitem[{{Franco-Hern{\'a}ndez} \&
  {Rodr{\'{\i}}guez}(2004)}]{Franco-HernandezRodriguez2004}
{Franco-Hern{\'a}ndez} R., {Rodr{\'{\i}}guez} L.~F., 2004, \apjl, 604, L105

\bibitem[{{Fryxell} {et~al}\mbox{.}(2000){Fryxell}, {Olson}, {Ricker},
  {Timmes}, {Zingale}, {Lamb}, {MacNeice}, {Rosner}, {Truran}, \&
  {Tufo}}]{Fryxell2000}
{Fryxell} B. {et~al.}, 2000, \apjs, 131, 273

\bibitem[{{Galv{\'a}n-Madrid} {et~al}\mbox{.}(2011){Galv{\'a}n-Madrid},
  {Peters}, {Keto}, {Mac Low}, {Banerjee}, \& {Klessen}}]{Galvan-Madrid+2011}
{Galv{\'a}n-Madrid} R., {Peters} T., {Keto} E.~R., {Mac Low} M.-M., {Banerjee}
  R., {Klessen} R.~S., 2011, \mnras, 1081

\bibitem[{{Galv{\'a}n-Madrid} {et~al}\mbox{.}(2008){Galv{\'a}n-Madrid},
  {Rodr{\'{\i}}guez}, {Ho}, \& {Keto}}]{Galvan-Madrid+2008}
{Galv{\'a}n-Madrid} R., {Rodr{\'{\i}}guez} L.~F., {Ho} P.~T.~P., {Keto} E.,
  2008, \apjl, 674, L33

\bibitem[{{G{\'o}mez} {et~al}\mbox{.}(2008){G{\'o}mez}, {Rodr{\'{\i}}guez},
  Loinard, Lizano, Allen, Poveda, \& Menten}]{gomezetal08}
{G{\'o}mez} L., {Rodr{\'{\i}}guez} L.~F., Loinard L., Lizano S., Allen C.,
  Poveda A., Menten K.~M., 2008, ApJ, 685, 333

\bibitem[{{Hoare} {et~al}\mbox{.}(2007){Hoare}, {Kurtz}, {Lizano}, {Keto}, \&
  {Hofner}}]{Hoare+2007}
{Hoare} M.~G., {Kurtz} S.~E., {Lizano} S., {Keto} E., {Hofner} P., 2007,
  Protostars and Planets V, 181

\bibitem[{{Hosokawa} \& {Omukai}(2009)}]{HosokawaOmukai2009}
{Hosokawa} T., {Omukai} K., 2009, \apj, 691, 823

\bibitem[{{Iliev} {et~al}\mbox{.}(2006){Iliev}, {Ciardi}, {Alvarez}, {Maselli},
  {Ferrara}, {Gnedin}, {Mellema}, {Nakamoto}, {Norman}, {Razoumov},
  {Rijkhorst}, {Ritzerveld}, {Shapiro}, {Susa}, {Umemura}, \&
  {Whalen}}]{Iliev2006}
{Iliev} I.~T. {et~al.}, 2006, \mnras, 371, 1057

\bibitem[{{Keto}(2002)}]{Keto2002}
{Keto} E., 2002, \apj, 580, 980

\bibitem[{{Keto}(2003)}]{Keto2003}
{Keto} E., 2003, \apj, 599, 1196

\bibitem[{{Keto}(2007)}]{Keto2007}
{Keto} E., 2007, \apj, 666, 976

\bibitem[{{Krumholz} {et~al}\mbox{.}(2010){Krumholz}, {Cunningham}, {Klein}, \&
  {McKee}}]{Krumholz+2010}
{Krumholz} M.~R., {Cunningham} A.~J., {Klein} R.~I., {McKee} C.~F., 2010, \apj,
  713, 1120

\bibitem[{{Krumholz} {et~al}\mbox{.}(2007){Krumholz}, {Klein}, \&
  {McKee}}]{Krumholz+2007}
{Krumholz} M.~R., {Klein} R.~I., {McKee} C.~F., 2007, \apj, 656, 959

\bibitem[{{Krumholz} {et~al}\mbox{.}(2011){Krumholz}, {Klein}, \&
  {McKee}}]{Krumholz+2011}
{Krumholz} M.~R., {Klein} R.~I., {McKee} C.~F., 2011, \apj, 740, 74

\bibitem[{{Krumholz} \& {Matzner}(2009)}]{KrumholzMatzner2009}
{Krumholz} M.~R., {Matzner} C.~D., 2009, \apj, 703, 1352

\bibitem[{{Matzner}(2002)}]{Matzner2002}
{Matzner} C.~D., 2002, \apj, 566, 302

\bibitem[{{McKee} \& {Tan}(2003)}]{McKeeTan2003}
{McKee} C.~F., {Tan} J.~C., 2003, \apj, 585, 850

\bibitem[{{Mezger} \& {Henderson}(1967)}]{MezgerHenderson1967}
{Mezger} P.~G., {Henderson} A.~P., 1967, \apj, 147, 471

\bibitem[{{Murray} {et~al}\mbox{.}(2011){Murray}, {M{\'e}nard}, \&
  {Thompson}}]{Murray+2011}
{Murray} N., {M{\'e}nard} B., {Thompson} T.~A., 2011, \apj, 735, 66

\bibitem[{{Nakano} {et~al}\mbox{.}(2000){Nakano}, {Hasegawa}, {Morino}, \&
  {Yamashita}}]{Nakano2000}
{Nakano} T., {Hasegawa} T., {Morino} J.-I., {Yamashita} T., 2000, \apj, 534,
  976

\bibitem[{{Nakano} {et~al}\mbox{.}(1995){Nakano}, {Hasegawa}, \&
  {Norman}}]{Nakano1995}
{Nakano} T., {Hasegawa} T., {Norman} C., 1995, \apss, 224, 523

\bibitem[{{Offner} {et~al}\mbox{.}(2009){Offner}, {Klein}, {McKee}, \&
  {Krumholz}}]{Offner2009}
{Offner} S.~S.~R., {Klein} R.~I., {McKee} C.~F., {Krumholz} M.~R., 2009, \apj,
  703, 131

\bibitem[{{Palla} \& {Stahler}(1991)}]{PallaStahler1991}
{Palla} F., {Stahler} S.~W., 1991, \apj, 375, 288

\bibitem[{{Palla} \& {Stahler}(1992)}]{PallaStahler1992}
{Palla} F., {Stahler} S.~W., 1992, \apj, 392, 667

\bibitem[{{Paxton}(2004)}]{Paxton2004}
{Paxton} B., 2004, \pasp, 116, 699

\bibitem[{{Peters} {et~al}\mbox{.}(2011){Peters}, {Banerjee}, {Klessen}, \&
  {Mac Low}}]{Peters2011}
{Peters} T., {Banerjee} R., {Klessen} R.~S., {Mac Low} M.-M., 2011, \apj, 729,
  72

\bibitem[{{Peters} {et~al}\mbox{.}(2010{\natexlab{a}}){Peters}, {Banerjee},
  {Klessen}, {Mac Low}, {Galv{\'a}n-Madrid}, \& {Keto}}]{Peters2010a}
{Peters} T., {Banerjee} R., {Klessen} R.~S., {Mac Low} M.-M.,
  {Galv{\'a}n-Madrid} R., {Keto} E.~R., 2010{\natexlab{a}}, \apj, 711, 1017

\bibitem[{{Peters} {et~al}\mbox{.}(2010{\natexlab{b}}){Peters}, {Klessen}, {Mac
  Low}, \& {Banerjee}}]{Peters2010c}
{Peters} T., {Klessen} R.~S., {Mac Low} M.-M., {Banerjee} R.,
  2010{\natexlab{b}}, \apj, 725, 134

\bibitem[{{Peters} {et~al}\mbox{.}(2010{\natexlab{c}}){Peters}, {Mac Low},
  {Banerjee}, {Klessen}, \& {Dullemond}}]{Peters2010b}
{Peters} T., {Mac Low} M.-M., {Banerjee} R., {Klessen} R.~S., {Dullemond}
  C.~P., 2010{\natexlab{c}}, \apj, 719, 831

\bibitem[{{Pollack} {et~al}\mbox{.}(1994){Pollack}, {Hollenbach}, {Beckwith},
  {Simonelli}, {Roush}, \& {Fong}}]{Pollack+2004}
{Pollack} J.~B., {Hollenbach} D., {Beckwith} S., {Simonelli} D.~P., {Roush} T.,
  {Fong} W., 1994, \apj, 421, 615

\bibitem[{{Rijkhorst} {et~al}\mbox{.}(2006){Rijkhorst}, {Plewa}, {Dubey}, \&
  {Mellema}}]{Rijkhorst}
{Rijkhorst} E.-J., {Plewa} T., {Dubey} A., {Mellema} G., 2006, \aap, 452, 907

\bibitem[{{Rodr{\'{\i}}guez} {et~al}\mbox{.}(2007){Rodr{\'{\i}}guez},
  {G{\'o}mez}, \& {Tafoya}}]{Rodriguez+2007}
{Rodr{\'{\i}}guez} L.~F., {G{\'o}mez} Y., {Tafoya} D., 2007, \apj, 663, 1083

\bibitem[{{Spitzer}(1978)}]{Spitzer1978}
{Spitzer} L., 1978, {Physical processes in the interstellar medium}. New York
  Wiley-Interscience, 1978.~333 p.

\bibitem[{{Tan} \& {McKee}(2004)}]{TanMcKee2004}
{Tan} J.~C., {McKee} C.~F., 2004, \apj, 603, 383

\bibitem[{{Testi} {et~al}\mbox{.}(2000){Testi}, {Sargent}, {Olmi}, \&
  {Onello}}]{Testi+2000}
{Testi} L., {Sargent} A.~I., {Olmi} L., {Onello} J.~S., 2000, \apjl, 540, L53

\bibitem[{{Tout} {et~al}\mbox{.}(1996){Tout}, {Pols}, {Eggleton}, \&
  {Han}}]{Tout1996}
{Tout} C.~A., {Pols} O.~R., {Eggleton} P.~P., {Han} Z., 1996, \mnras, 281, 257

\bibitem[{{Williams} {et~al}\mbox{.}(2000){Williams}, {Blitz}, \&
  {McKee}}]{Williams+2000}
{Williams} J.~P., {Blitz} L., {McKee} C.~F., 2000, Protostars and Planets IV,
  97

\bibitem[{{Wood} \& {Churchwell}(1989)}]{WoodChurchwell1989}
{Wood} D.~O.~S., {Churchwell} E., 1989, \apjs, 69, 831

\bibitem[{{Yorke} \& {Sonnhalter}(2002)}]{YorkeSonnhalter2002}
{Yorke} H.~W., {Sonnhalter} C., 2002, \apj, 569, 846

\bibitem[{{Zinnecker} \& {Yorke}(2007)}]{ZinneckerYorke2007}
{Zinnecker} H., {Yorke} H.~W., 2007, \araa, 45, 481

\end{thebibliography}

\label{lastpage}

\end{document}